\documentclass[11pt]{article}
\usepackage{amsmath,amssymb,color,graphics,epsfig}

\usepackage{amsfonts}

\newcommand{\hoch}[1]{$\, ^{#1}$}

\makeatletter
\@addtoreset{equation}{section}
\makeatother

\newcommand{\be}{\begin{equation}}
\newcommand{\ee}{\end{equation}}
\newcommand{\bea}{\setlength\arraycolsep{2pt} \begin{eqnarray}}
\newcommand{\eea}{\end{eqnarray}}
\newcommand{\nn}{\nonumber}

\def\ft#1#2{{\textstyle{\frac{\scriptstyle #1}{\scriptstyle #2} } }}
\def\fft#1#2{{\frac{#1}{#2}}}

\def\0{{\sst{(0)}}}
\def\1{{\sst{(1)}}}
\def\2{{\sst{(2)}}}
\def\3{{\sst{(3)}}}
\def\4{{\sst{(4)}}}
\def\5{{\sst{(5)}}}
\def\6{{\sst{(6)}}}
\def\7{{\sst{(7)}}}
\def\8{{\sst{(8)}}}
\def\sst#1{{\scriptscriptstyle #1}}

\begin{document}

\begin{flushright}
\end{flushright}

\vspace{25pt}
\begin{center}
{\large {\bf Axial Gravitational Perturbations of Slowly-Rotating Compact Objects in General Relativity and Beyond}}

\vspace{40pt}
{\bf Xing-Hui Feng\hoch{1*} and Jun Peng\hoch{2\dagger}}

\vspace{10pt}

\hoch{1}{\it Center for Joint Quantum Studies and Department of Physics,
School of Science, Tianjin University, Tianjin 300350, China}

\hoch{2}{\it Institute of Theoretical Physics, School of Physics and Optoelectronic Engineering, Beijing University of Technology, Beijing 100124, China}

\vspace{40pt}

\underline{ABSTRACT}
\end{center}

We study the axial gravitational perturbations of slowly-rotating compact objects which are assumed to be supported by anisotropic fluids. We find that the gravitational perturbations decouple from the matter perturbations for axial sectors. We obtain a master wave equation whose potential is fully determined by the metric functions. This equation makes the calculations of gravitational QNMs for rotating compact objects extremely easy in specific background configurations.

\vfill {\footnotesize \hoch{*} xhfeng@tju.edu.cn  \ \ \hoch{\dagger} junpeng@bjut.edu.cn}

\thispagestyle{empty}

\pagebreak

\addtocontents{toc}{\protect\setcounter{tocdepth}{2}}


\vspace{1cm}

\section{Introduction}
Black hole (BH) perturbation theory is an important tool to study the stability and quasi-normal modes (QNMs) of BH, the ringdown signal from compact binary coalescences \cite{Kokkotas:1999bd,Berti:2009kk,Konoplya:2011qq}. The metric perturbations of static BHs can be well established according to their spherical symmetry. We can expand the metric perturbations using spherical harmonics and separate them into the axial and polar parts depending on their behavior under parity transformations. The axial perturbations of Schwarzschild BH is studied in \cite{Regge:1957td} and the master wave equation is the famous Regge-Wheeler equation. The polar perturbations of Schwarzschild BH is studied in \cite{Zerilli:1970wzz,Zerilli:1970se} and the master wave equation is the so-called Zerilli equation.

However the harmonics decomposition of metric perturbations is not suitable for rotating BH. The perturbations of Kerr BH are completed in \cite{Teukolsky:1973ha} based on Newman-Penrose (NP) formalism. The master wave equation is the famous Teukolsky equation, whose solutions are two-dimensional eigenvalue problem. When applying the NP form to the perturbations of Kerr-Newman (KN) BH, we have no ways to obtain a variables-separated and perturbations-decoupled master wave equation \cite{Chandrasekhar:1985kt}. Anyway the analytical perturbations of non-Kerr and spinning BH in (or beyond) general relativity (GR) are open problems. It should be point out that one can obtain modified Teukolsky equations when we consider the deviations from Kerr BH is perturbative \cite{Li:2022pcy,Cano:2023tmv,Cano:2023jbk}.

Things become possible when we consider the slowly-rotating backgrounds. Since the spin parameter is viewed as a small modification with respect to the static metric, the general decomposition of metric perturbations, as well as the gauge choosing are still valid for slowly-rotating ones. Based on Kojima's elegant work about the angular separation of perturbation equations \cite{Kojima:1992ie}, Pani etal proposed an advanced perturbation method for slowly-rotating spacetime \cite{Pani:2012bp,Brito:2013wya,Pani:2013pma}. This advanced method is applied in the perturbations of Kerr BH \cite{Pani:2013pma}, KN BH \cite{Pani:2013ija,Pani:2013wsa}, BHs in Einstein-dilaton Gauss-Bonnet gravity \cite{Pierini:2021jxd,Pierini:2022eim}, dynamical Chern-Simons gravity \cite{Wagle:2021tam,Srivastava:2021imr}, higher-derivative gravity \cite{Cano:2021myl}, Einstein-Bumblebee gravity \cite{Liu:2022dcn}.

In this paper, we investigate the axial gravitational perturbations of slowly-rotating compact objects in GR and beyond. We assume the general slowly-rotating metric ansatz is supported by the effective stress tensor of an anisotropic fluid, seeing the set up in section 2. We analyse the axial gravitational perturbations in section 3 and derive a master wave equation in section 4. We obtain a parametrized QNMs expression of KN BH by integrating the master wave equation in section 5. We give a brief discussion in section 6.

\section{The set up}
Understanding the gravitational field of a rotating body is one of great astrophysical and theoretical significance. Although a number of exact solutions with rotation are known in Einstein gravity, for general modified gravities they are quite difficult to obtain.  On the other hand slowly-rotating solutions are somewhat easier to
come by. The general ansatz of slowly-rotating metric is
\be
\bar{ds}^2=-A(r)dt^2+\frac{dr^2}{B(r)}+C(r)(d\theta^2+\sin^2\theta d\phi^2)-2aD(r)\sin^2\theta dtd\phi\label{srmetric}
\ee
where $a$ is the spin parameter. In ab-initio calculation, the metric should solve the equations of motion for certain theory of gravity to leading order in the spin parameter $a$. From a phenomenological point of view, we can assume the solution is governed by the Einstein equation with an effective stress tensor
\be
R_{\mu\nu}-\fft12g_{\mu\nu}R=8\pi T_{\mu\nu}
\ee
Without loss of generality, the stress tensor supporting the metric \eqref{srmetric} is described by an anisotropic fluid \cite{Raposo:2018rjn}
\be
T_{\mu\nu}=\rho u_\mu u_\nu + P_r k_\mu k_\nu + P_t\Pi_{\mu\nu}
\ee
where $u_\mu$ is the fluid's 4-velocity and $k_\mu$ is the unit spacelike vector orthogonal to $u_\mu$, i.e.
\be
-u^\mu u_\mu=k^\mu k_\mu=1,\quad u^\mu k_\mu=0 \label{conditions}
\ee
and
\be
\Pi_{\mu\nu}=g_{\mu\nu}+u_\mu u_\nu-k_\mu k_\nu
\ee
is a projection operator onto a two-surface orthogonal to $u^\mu$ and $k^\mu$, i.e.
\be
u_\mu \Pi^{\mu\nu} V_\nu=k_\mu \Pi^{\mu\nu} V_\nu=0
\ee
for any vector $V^\mu$. We can assume
\be
u^\mu=\frac{1}{\sqrt A}(1,0,0,\Omega),\quad k^\mu=\sqrt B(0,1,0,0)
\ee
where $\Omega$ is the fluid angular velocity.

\section{Axial gravitational perturbations}
In order to study the gravitational perturbations, one needs to perturb the gravitational equations as well as the energy momentum tensor. As in the spherically symmetric case, the metric perturbations $g_{\mu\nu}=\bar g_{\mu\nu}+h_{\mu\nu}$ can be decomposed into the axial and polar parts. In Regge-Wheeler gauge they reads
\bea
h^{-}_{\mu\nu}dx^\mu dx^\nu&=&2[h_0^{\ell m}dt+h_1^{\ell m}dr]\times[S_\theta^{\ell m}d\theta+S_\phi^{\ell m}d\phi]\nn\\
h^{+}_{\mu\nu}dx^\mu dx^\nu&=&[AH_0^{\ell m}dt^2+2H_1^{\ell m}dtdr+H_2^{\ell m}/Bdr^2+K^{\ell m}d\Omega_2^2]Y^{\ell m}\label{metricpert}
\eea
where $h_0^{\ell m},h_1^{\ell m},H_0^{\ell m},H_1^{\ell m},H_2^{\ell m},K^{\ell m}$ are free functions of $t, r$, and $S_a^{\ell m}$ is the axial vector spherical harmonics
\be
(S_\theta^{\ell m},S_\phi^{\ell m})\equiv\left(-\frac{Y^{\ell m}_{,\phi}}{\sin\theta},\sin\theta Y^{\ell m}_{,\theta}\right)
\ee
In the case of a spherically symmetric background, perturbations with different values of $(\ell,m)$, as well as perturbations with opposite parity, are decoupled. In a rotating, axially symmetric background, perturbations with different values of $m$ are still decoupled but perturbations with different values of $\ell$ are not. However in \cite{Pani:2012bp,kojima1993}, it's proved that if we are only interested in the eigenvalue spectrum to first order in spin $a$, the polar and axial perturbations--as well as perturbations with different values of the harmonic indices--are decoupled from each other, and can be studied independently. In the following, we only consider axial perturbations and continue in tetrad formalism for simplicity.
We define the tetrad basis $e^{\bar a}=e^{\bar a}_\mu dx^\mu$ as
\bea
e^{\bar0}&=&\sqrt{A}dt+\frac{aD\sin^2\theta}{\sqrt{A}}d\phi,\quad e^{\bar1}=\frac{dr}{\sqrt{B}},\nn\\ e^{\bar2}&=&\sqrt{C}d\theta+\frac{h_0^{\ell m}S_\theta^{\ell m}}{\sqrt{C}}dt+\frac{h_1^{\ell m}S_\theta^{\ell m}}{\sqrt{C}}dr\nn\\
e^{\bar3}&=&\sqrt{C}\sin\theta d\phi+\frac{h_0^{\ell m}S_\phi^{\ell m}}{\sqrt{C}\sin\theta}dt+\frac{h_1^{\ell m}S_\phi^{\ell m}}{\sqrt{C}\sin\theta}dr
\eea
The inverse tetrad $E^\mu_{\bar a}$ are given by
\bea
e^{\bar a}_\mu E^\nu_{\bar a}=\delta_\mu^\nu,\quad e^{\bar a}_\mu E^\mu_{\bar b}=\delta^{\bar a}_{\bar b}
\eea
The tensor components in coordinate and tetrad basis are related by
\bea
A_\mu&=&e^{\bar a}_\mu A_{\bar a},\quad A_{\bar a}=E^\mu_{\bar a}A_\mu\nn\\
B_{\mu\nu}&=&e^{\bar a}_\mu e^{\bar b}_\nu B_{\bar a\bar b},\quad B_{\bar a\bar b}=E^\mu_{\bar a}E^\nu_{\bar b}B_{\mu\nu}
\eea
In tetrad formalism, we rewrite the effective Einstein equation as
\be
R_{\bar a\bar b}=8\pi T^{t}_{\bar a\bar b}
\ee
where
\be
T^t_{\bar a\bar b}\equiv\left(T_{\bar a\bar b}-\ft12T\eta_{\bar a\bar b}\right)
\ee
The perturbation equations are given by
\be
{\cal E}_{\bar a\bar b}\equiv\delta R_{\bar a\bar b}-8\pi\delta T^t_{\bar a\bar b}=0
\ee
The perturbations of Ricci tensor can be calculated based on the metric perturbation \eqref{metricpert}. The perturbations of stress tensor are discussed in appendix \ref{appA} which involves two axial variables $U^{\ell m}$ and $Q^{\ell m}$. The axial perturbation equations can be divided into two groups \cite{Kojima:1992ie,Pani:2013pma}. The first group is
\bea
{\cal E}_{\bar L\bar2}&=&a\tilde\alpha^L_{\ell m}\cos\theta Y^{\ell m}_{,\theta}-\beta^L_{\ell m}\frac{Y^{\ell m}_{,\phi}}{\sin\theta}+a\eta^L_{\ell m}\sin\theta Y^{\ell m}+a\chi^L_{\ell m}\sin\theta W^{\ell m}\nn\\
{\cal E}_{\bar L\bar3}&=&\beta^L_{\ell m} Y^{\ell m}_{,\theta}+a\tilde\alpha^L_{\ell m}\cos\theta\frac{Y^{\ell m}_{,\phi}}{\sin\theta}+a\chi^L_{\ell m} X^{\ell m}
\eea
where $L=0, 1$, and
\be
(X^{\ell m},W^{\ell m})\equiv\left(2(Y^{\ell m}_{,\theta\phi}-\cot\theta Y^{\ell m}_{,\phi}),Y^{\ell m}_{,\theta\theta}-\cot\theta Y^{\ell m}_{,\theta}-\frac{Y^{\ell m}_{,\phi\phi}}{\sin^2\theta}\right)
\ee
span the rank-two tensor harmonics. The second group reads
\bea
{\cal E}_{\bar2\bar3}&=&ag_{\ell m}Y^{\ell m}_{,\phi}+t_{\ell m}W^{\ell m}\nn\\
{\cal E}_{(-)}={\cal E}_{\bar2\bar2}-{\cal E}_{\bar3\bar3}&=&ag_{\ell m}\sin\theta Y^{\ell m}_{,\theta}-t_{\ell m}\frac{X^{\ell m}}{\sin\theta}
\eea
Here $\tilde\alpha^L_{\ell m},\beta^L_{\ell m},\eta^L_{\ell m},\chi^L_{\ell m},t_{\ell m},g_{\ell m}$ consist of axial perturbation variables $h_0^{\ell m},h_1^{\ell m},U^{\ell m},Q^{\ell m}$. Their explicit expressions are given in appendix \ref{appB}, where all perturbation variables are in frequency domain with Fourier transformation $e^{-i\omega t}$.

\section{Master wave equation}
In order to derive a master wave equation, we need separate the angular dependence. This can be performed using the orthogonality properties of the spherical harmonics \cite{Kojima:1992ie,Pani:2013pma}
\bea
P_L&=&\int d\Omega\left[Y^{\ast\ell'm'}_{,\theta}{\cal E}_{\bar L\bar3}-\frac{Y^{\ast\ell'm'}_{,\phi}}{\sin\theta}{\cal E}_{\bar L\bar2}\right]\nn\\
&=&J\beta^L_{\ell m}+ima[(J-2)\chi^L_{\ell m}+\tilde\alpha^L_{\ell m}+\eta^L_{\ell m}]=0\\
P_2&=&\int d\Omega\frac{1}{J-2}\left[W^{\ast\ell'm'}{\cal E}_{\bar2\bar3}-\frac{X^{\ast\ell'm'}}{\sin\theta}{\cal E}_{(-)}\right]\nn\\
&=&Jt_{\ell m}+imag_{\ell m}=0
\eea
where $d\Omega=\sin\theta d\theta d\phi$ is the volume form of the two-sphere and $J=\ell(\ell+1)$. We have three equations $P_0=P_1=P_2=0$ for the axial perturbations $h_0^{\ell m}, h_1^{\ell m}$ and $U^{\ell m}$, in which
\bea
P_1&=&J\beta^1_{\ell m}+ima[(J-2)\chi^1_{\ell m}+\tilde\alpha^1_{\ell m}+\eta^1_{\ell m}]=0\\
P_2&=&Jt_{\ell m}+imag_{\ell m}=0
\eea
are two first order equations involved only in $h_0^{\ell m}, h_1^{\ell m}$. We start with the perturbation of static backgroumd
\bea
\beta^1_{\ell m}&=&0\nn\\
t_{\ell m}&=&0
\eea
The second equation gives
\be
h_0=a(r)h_1+b(r)h_1'
\ee
From now on, we omit the upscript $\ell m$ in all perturbation variables. Substituting it into the first equation, we have
\be
h_1''+c(r)h_1'+d(r)h_1=0
\ee
where $a(r), b(r), c(r), d(r)$ are certain functions. Finally, we can introduce a master variable
\be
\Psi^{(0)}=\sqrt{\frac{AB}{C}}h_1
\ee
Then we obtain a Schrodinger-like wave equation
\be
\frac{d^2\Psi^{(0)}}{dr_\ast^2}+(\omega^2-V^{(0)})\Psi=0
\ee
where $r_\ast$ is the tortoise coordinate defined by $dr_\ast=dr/\sqrt{AB}$. The potential $V$ is given by
\be
V^{(0)}=(J-2)\frac{A}{C}+\sqrt{ABC}\left(\sqrt{AB}\left(\frac{1}{\sqrt{C}}\right)'\right)'
\ee
Note that the equation $P_0=0$ gives the solution of fluid perturbation $U^{\ell m}$
\be
U^{\ell m}=-h_0^{\ell m}\label{solutionU}
\ee
Now we turn to the rotating background. Since we consider only the slowly-rotating case, i.e. the spin parameter $a$ contributes a linear correction to the static case, we can assume the master variable has the following form
\be
\Psi=\sqrt{\frac{AB}{C}}(h_1+maf_{h_1})
\ee
Then we can also obtain a Schrodinger-like wave equation
\be
\frac{d^2\Psi}{dr_\ast^2}+(\omega^2-V^{(0)}-maV^{(a)})\Psi=0\label{waveeq}
\ee
We find that the corrective function $f_{h_1}$ is
\be
f_{h_1}=c+\frac1\omega\frac{D}{C}
\ee
where $c$ is a constant and the corrective potential $V^{(a)}$ is
\be
V^{(a)}=2\omega\frac{D}{C}+\frac{1}{J\omega}\left[\sqrt{AB}\left(\sqrt{AB}\left(\frac{D}{C}\right)'\right)'-AB\frac{C'}{C}\left(\frac{D}{C}\right)'\right]
\ee
We can see that the axial gravitational perturbations decouple from perturbations of the effective stress tensor. A similar result was derived in \cite{Hui:2021cpm} to parameterize deviations from general relativistic expectations at the level of the action governing the dynamics of BH perturbations. Their result applies to perturbations around BHs with scalar hair. In fact, our parameterized deviations attribute to the phenomenological stress tensor. Thus our result is more applicable, even for theories with vector hairs and high-dimensional operators whose perturbation dynamics is neglected. Another important playground of the master wave equation \eqref{waveeq} is galactic BHs immersed in some environment, like an accretion disk or dark matter halo. It can be exploited for quick and quantitative estimates of QNM excitation in astrophysical scenarios \cite{Barausse:2014tra}. It is also interesting to extend our analysis to self-gravitating anisotropic compact stars \cite{Kumar:2022kho} and investigate their tidal deformability \cite{Pani:2015nua,Biswas:2019gkw}.

\section{Parametrized QNMs of Kerr-Newman black hole}

\begin{center}
\begin{figure}
  \includegraphics[width=7cm]{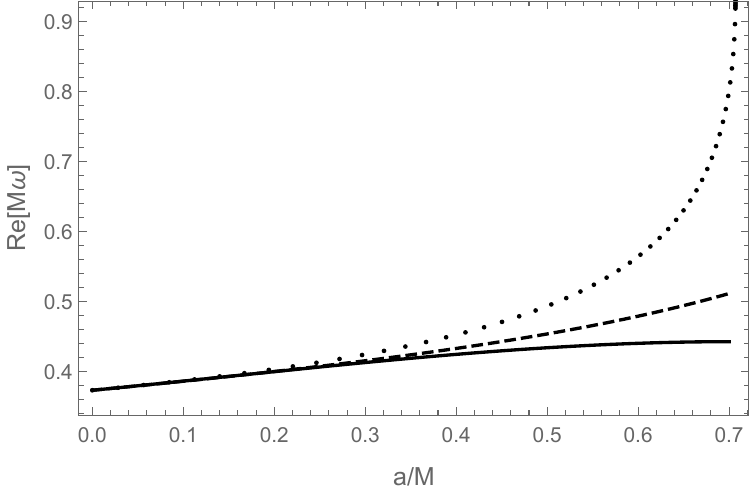}\qquad\includegraphics[width=7cm]{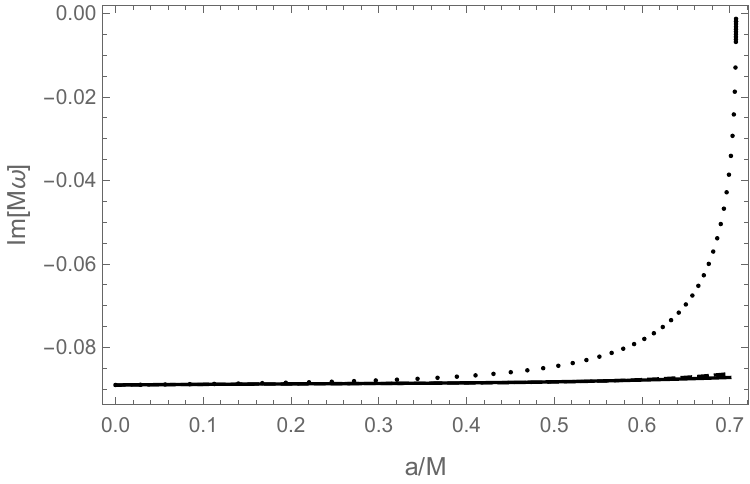}\\
  \caption{The fundamental gravitational QNMs of KN BH with $Q=a$ for $\ell=m=2$. Dotted, dashed, solid lines denote exact numerical calculation in \cite{Dias:2015wqa}, slow-rotation approximation in \cite{Pani:2013ija,Pani:2013wsa} and parameterized QNMs expression \eqref{paraqnm}, respectively.}\label{KNQNMs}
\end{figure}
\end{center}

In this section, as an example of application, we use the master wave equation \eqref{waveeq} to derive a parameterized QNMs expression of KN BH. At first order in spin $a$ for KN BH, we have
\be
A(r)=B(r)=\Delta/r^2,\quad C(r)=r^2,\quad D(r)=\frac{2M}{r}-\frac{Q^2}{r^2}
\ee
where $\Delta=r^2-2Mr+Q^2$, the potentials reduce to
\bea
V^{(0)}&=&\frac{\Delta}{r^5}\left[Jr-6M+\frac{4Q^2}{r}\right]\nn\\
V^{(a)}&=&\frac{8 \Delta \left(3 M r^2 (3r-7 M)+7 Q^2 r (4M-r)-9 Q^4\right)}{J\omega r^{10}}+\frac{2\omega \left(2Mr-Q^2\right)}{r^4}
\eea
Since we neglect the perturbations of electromagnetic field, these results suppose to give the gravitational QNMs of slowly-rotating and weakly-charged KN BH. To confirm this, we apply the parametrized QNMs formalism developed in \cite{Cardoso:2019mqo} to the master wave equation \eqref{waveeq}. At first order of $a$, we rewrite the equation \eqref{waveeq} as
\be
Zf\frac{d}{dr}\left(Zf\frac{d\Psi}{dr}\right)+\left[\left(\omega-\frac{maD}{r^2}\right)^2-ZfV_\Psi\right]\Psi=0\label{eqPsi}
\ee
with
\bea
f(r)&=&1-\frac{r_H}{r},\quad Z=1-\frac{r_-}{r},\quad D=\frac{r_H+r_-}{r}-\frac{r_Hr_-}{r^2}\\
V_\Psi&=&\frac{\ell(\ell+1)}{r^2}-\frac{3(r_H+r_-)}{r^3}+\frac{4r_Hr_-}{r^4}+\frac{ma}{\ell(\ell+1)\omega}\left[\frac{36(r_H+r_-)}{r^5}\right.\nn\\
&&\left.-\frac{14(3r_H+r_-)(r_H+3r_-)}{r^6}+\frac{112r_Hr_-(r_H+r_-)}{r^7}-\frac{72r_H^2r_-^2}{r^8}\right]
\eea
Introducing a new master variable $\Phi=\sqrt Z\Psi$, then the equation \eqref{eqPsi} becomes
\be
f\frac{d}{dr}\left(f\frac{d\Phi}{dr}\right)+\left[\frac{1}{Z^2}\left(\omega-\frac{maD}{r^2}\right)^2-fV_\Phi\right]\Phi=0\label{eqPhi}
\ee
with
\be
V_\Phi=\frac{V_\Psi}{Z}-\frac{f(Z')^2-2Z(fZ')'}{4Z^2}
\ee
At first order of $r_-$, we can rewrite the equation \eqref{eqPhi} in the form
\be
f\frac{d}{dr}\left(f\frac{d\Phi}{dr}\right)+\left[\left(1-\frac{r_-}{r_H}\right)^{-2}\left(\omega-\frac{ma}{r_H^2}\right)^2-f(V_-+\delta V)\right]\Phi=0
\ee
with
\bea
V_- &=& \frac{\ell(\ell+1)}{r^2}-\frac{3r_H}{r^3}\nn\\
\delta V &=& \frac{1}{r_H^2}\sum^7_{j=0}\alpha_j^-\left(\frac{r_H}{r}\right)^j
\eea
We obtain eight terms in the expansion of $\delta V$, and coefficients $\alpha_j^-$ can be read off straightforward. The parametrized QNMs are given by
\be
\omega=\left(1-\frac{r_-}{r_H}\right)\frac{\Omega_0}{r_H}+\frac{ma}{r_H^2}+\sum_{j=0}^7\alpha_j^-e_j^-
\ee
For the fundamental mode with $\ell=2$, the QNM frequency at ${\cal O}(a,Q^2)$ becomes
\bea
M\omega&=&0.3736717-0.0889623i+\frac{(0.0648115-0.0053026i)Q^2}{M^2}\nn\\
&&+\frac{am}{M}\left(0.0628831 + 0.0009979i-\frac{(0.0734411 - 0.0043788i)Q^2}{M^2}\right)\label{paraqnm}
\eea
where we used $r_H=M+\sqrt{M^2-Q^2}$, $r_-=M-\sqrt{M^2-Q^2}$ and the numerical data of $e_j^-$ \cite{Cardoso:2019mqo}. When $Q=0$, it reduces to the parameterized QNMs of slowly-rotating Kerr BH obtained in \cite{Hatsuda:2020egs}. We can see from Figure \ref{KNQNMs} that our parametrized expression \eqref{paraqnm} gives QNMs of KN BH in excellent agreement with numerical results reported in \cite{Dias:2015wqa} when $a$ and $Q$ are small.

\section{Discussion}
Gravitational waves emitted during the ringdown phase are determined by QNMs and offer valuable insight into the nature of the objects. The frequency and damping of these oscillations depend only on the parameters characterizing the BH. Thus tests for GR prediction and possible deviations are very important. In this paper, we introduce a phenomenological fluid to parametrize deviations from GR at the level of equations of motion and study the axial gravitational perturbations of general slowly-rotating compact objects. We obtain a modified Regge-Wheeler equation whose potential is fully determined by the metric functions. This equation makes the calculations of gravitational QNMs for rotating compact objects extremely easy in specific background configurations. We leave the polar perturbations for future's work.

\section*{Acknowledgement}
X.H.F. is supported by NSFC (National Natural Science Foundation of China) Grant No. 11905157 and No. 11935009.

\appendix
\section{Perturbations of stress tensor}\label{appA}
The most general perturbations of the stress tensor are
\bea
\delta T^{\mu\nu}=(\rho+P_t)\delta(u^\mu u^\nu)+(P_r-P_t)\delta(k^\mu k^\nu)+P_t\delta g^{\mu\nu}\nn\\
+(\delta\rho+\delta P_t)u^\mu u^\nu+(\delta P_r-\delta P_t)k^\mu k^\nu+\delta P_t g^{\mu\nu}
\eea
The perturbation of the energy density and pressures can be expanded by the scalar spherical harmonics
\be
\delta\rho=\delta\rho^{\ell m}Y^{\ell m},\quad\delta P_r=\delta P_r^{\ell m}Y^{\ell m},\quad\delta P_t=\delta P_t^{\ell m}Y^{\ell m}
\ee
Similar to the decomposition of metric perturbation, the perturbation of fluid's 4-velocity $u^\mu$ are described by three functions $R^{\ell m},U^{\ell m},V^{\ell m}$
\bea
\delta u^r &=& \frac{1}{\sqrt AB}R^{\ell m}Y^{\ell m}\nn\\
\delta u^\theta &=& \frac{1}{\sqrt AC}(U^{\ell m}S^{\ell m}_\theta+V^{\ell m}Y^{\ell m}_{,\theta})\nn\\
\delta u^\phi &=& \frac{1}{\sqrt AC\sin^2\theta}(U^{\ell m}S^{\ell m}_\phi+V^{\ell m}Y^{\ell m}_{,\phi})
\eea
The perturbations of $k^\mu$ are described by two functions $Q^{\ell m}, P^{\ell m}$
\bea
\delta k^\theta&=&\frac{\sqrt B}{C}(Q^{\ell m}S^{\ell m}_\theta+P^{\ell m}Y^{\ell m}_{,\theta})\nn\\
\delta k^\phi&=&\frac{\sqrt B}{C\sin^2\theta}(Q^{\ell m}S^{\ell m}_\phi+P^{\ell m}Y^{\ell m}_{,\phi})
\eea
The other perturbed components are determined by conditions \eqref{conditions}. According to the classifications, $R^{\ell m},V^{\ell m},P^{\ell m},\delta\rho^{\ell m},\delta P_r^{\ell m},\delta P_t^{\ell m}$ belong to the polar perturbations, and $U^{\ell m}, Q^{\ell m}$ belong to axial perturbations. In this paper, we only consider the axial perturbations.  Eventually, we obtain the components of perturbed stress tensor in tetrad form
\be
\delta T^t_{\bar a\bar b}=\left(
                \begin{array}{cccc}
                  \frac{2a\Omega\kappa(U^{\ell m}+h_0^{\ell m})S_\phi^{\ell m}}{A} & -\frac{a\Omega\sqrt B\sigma(Q^{\ell m}+h_1^{\ell m})S_\phi^{\ell m}}{\sqrt A} & -\frac{\sigma(U^{\ell m}+h_0^{\ell m})S_\theta^{\ell m}}{\sqrt{AC}} & -\frac{\sigma(U^{\ell m}+h_0^{\ell m})S_\phi^{\ell m}}{\sqrt{AC}\sin\theta} \\
                  sym & 0 & \frac{\sqrt B\sigma(Q^{\ell m}+h_1^{\ell m})S_\theta^{\ell m}}{\sqrt C} & \frac{\sqrt B\sigma(Q^{\ell m}+h_1^{\ell m})S_\phi^{\ell m}}{\sqrt C\sin\theta} \\
                  sym & sym & 0 & \frac{a\Omega\kappa(U^{\ell m}+h_0^{\ell m})\sin\theta S_\theta^{\ell m}}{A} \\
                  sym & sym & sym & \frac{2a\Omega\kappa(U^{\ell m}+h_0^{\ell m})S_\phi^{\ell m}}{A} \\
                \end{array}
              \right)\label{pertST}
\ee
where we have defined $\kappa=\rho+P_t$ and $\sigma=P_r-P_t$. We can see from the perturbed stress tensor \ref{pertST} that $Q^{\ell m}+h_1^{\ell m}$ corresponds to the momentum flow. Since the fluid we consider has no friction, we set
\be
Q^{\ell m}+h_1^{\ell m}=0
\ee

\section{Axial perturbation variables}\label{appB}
\bea
\beta^0&=&\frac{\left(-B A' C'+A B' C'+2 A \left(B C''+J-2\right)\right)h_0}{4 (AC)^{3/2}}+\frac{\left(B A'-A B'\right)h_0'}{4 A^{3/2} \sqrt{C}}-\frac{B h_0''}{2 \sqrt{AC}}\nn\\
&&-\frac{i \omega \left(-B C A'+A C B'+2 A B C'\right)h_1}{4 (AC)^{3/2}}-\frac{i \omega B h_1'}{2 \sqrt{AC}}+\frac{8\pi\kappa(U+h_0)}{\sqrt{AC}}\nn\\
\beta^1&=&-\frac{i \omega \sqrt{B} C' h_0}{2 A C^{3/2}}+\frac{i \omega \sqrt{B} h_0'}{2 A \sqrt{C}}+\left(\frac{\sqrt{B} (J-2)}{2 C^{3/2}}-\frac{\sqrt{B} \omega^2}{2 A \sqrt{C}}\right)h_1-\frac{8\pi\sqrt B\sigma(Q+h_1)}{\sqrt C}\nn\\
\chi^0&=&-\frac{i \omega D h_0 }{2 (AC)^{3/2}}+\frac{\left(A B D'-B D A'\right)h_1}{4 (AC)^{3/2}}\nn\\
\chi^1&=&-\frac{\sqrt{B} \left(C D A'+A D C'-A C D'\right)h_0}{4 A^2 C^{5/2}}+\frac{\sqrt{B} D h_0'}{4 A C^{3/2}}-\frac{i \omega \sqrt{B} D h_1}{4 A C^{3/2}}\nn\\
\tilde\alpha^0&=&\frac{\left(C D B'-2 B D C'+3 B C D'\right)h_1}{2 \sqrt{A} C^{5/2}}+\frac{B D h_1'}{\sqrt{A} C^{3/2}}\nn\\
\tilde\alpha^1&=&-\frac{\sqrt{B} \left(C D A'+3 A D C'-3 A C D'\right)h_0}{2 A^2 C^{5/2}}+\frac{\sqrt{B} D h_0'}{2 A C^{3/2}}-\frac{i \omega \sqrt{B} D h_1}{2 A C^{3/2}}\nn\\
\eta^0&=&\frac{J B \left(A D'-D A'\right)h_1}{4 (AC)^{3/2}}\nn\\
\eta^1&=&\frac{J \sqrt{B} \left(C D A'+3 A D C'-A C D'\right)h_0}{4 A^2 C^{5/2}}-\frac{3 J \sqrt{B} D  h_0'}{4 A C^{3/2}}-\frac{3 J i \omega \sqrt{B} D h_1}{4 A C^{3/2}}\nn\\
t&=&\frac{i \omega h_0}{2 A C}+\frac{\left(B A'+A B'\right)h_1}{4 A C}+\frac{B h_1'}{2 C}\nn\\
g&=&\frac{\left(-B C D A' C'+A C D \left(B' C'+2 B C''-4\right)+2 A B C C' D'-2 A B D C'^2\right)h_0}{4 A^2 C^3}\nn\\
&&+\frac{\left(B C D A'-A C D B'+2 A B D C'-2 A B C D'\right)h_0'}{4 A^2 C^2}-\frac{B D h_0''}{2 A C}\nn\\
&&-\frac{i \omega \left(-B D A'+A D B'+2 A B D'\right)h_1}{4 A^2 C}-\frac{i \omega B D  h_1'}{2 A C}+\frac{8\pi\Omega\kappa(U+h_0)}{A}\nn\\
\eea


\end{document}